# Dual Graph-Laplacian PCA: A Closed-Form Solution for Bi-clustering to Find "Checkerboard" Structures on Gene Expression Data

Jin-Xing Liu, *Member, IEEE*, Chun-Mei Feng, Xiang-Zhen Kong, Yong Xu, *Senior Member, IEEE*

**Abstract**—In the context of cancer, internal "checkerboard" structures are normally found in the matrices of gene expression data, which correspond to genes that are significantly up- or down-regulated in patients with specific types of tumors. In this paper, we propose a novel method, called dual graph-regularization principal component analysis (DGPCA). The main innovation of this method is that it simultaneously considers the internal geometric structures of the condition manifold and the gene manifold. Specifically, we obtain principal components (PCs) to represent the data and approximate the cluster membership indicators through Laplacian embedding. This new method is endowed with internal geometric structures, such as the condition manifold and gene manifold, which are both suitable for bi-clustering. A closed-form solution is provided for DGPCA. We apply this new method to simultaneously cluster genes and conditions (e.g., different samples) with the aim of finding internal "checkerboard" structures on gene expression data, if they exist. Then, we use this new method to identify regulatory genes under the particular conditions and to compare the results with those of other state-of-the-art PCA-based methods. Promising results on gene expression data have been verified by extensive experiments.

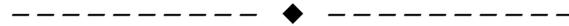

## 1 INTRODUCTION

### 1.1 Biological analysis of PCA

With the development of molecular biology, the gene chip has become one of the most important technologies of gene functional annotation in the post-genomic era [1]. Determining how to excavate reliable information from the high-throughput and multivariable gene chip to explain the regulatory network of gene function is the bottleneck problem of bioinformatics [2]. Without losing the original data, principal component analysis (PCA) transforms the data to a low-dimensional linear or nearly linear subspace constituted by principal components (PCs) [3]. This method overcomes the limitations of bioinformatics methods in gene chip analysis and provides new inspiration for biological data mining. For example, the selected information simplifies the complexity of the gene chip variable and clusters the obtained data. This method provides the basis for early diagnosis and subtyping of cancer.

### 1.2 Checkerboard structures in gene expression data and relations with PCA

In the absence of class knowledge of genes and samples, it is necessary to find potential classes by exploiting the relationship between genes and conditions. Bi-clustering exploits the potential two-sided data structure, which helps the two-dimensional clustering make meaningful study of genes and samples. This method has achieved better results than using single-dimensional clustering to cluster conditions or features independently [4-7]. Since the gene expression data generated by gene chip technology are expressed as a "high-dimensional small-sample" matrix, we assume that there are "checkerboard" structures in these data, which is reasonable and effective [8]. Bi-clustering can be used to find the checkerboard structures hidden in a gene expression data matrix, which has been well studied [8]. Specifically, bi-clustering is performed in both the row and column directions simultaneously, which interact with and restrain each other, to identify the checkerboard structures within the gene expression data, if they exist. These checkerboard structures are formed by the network of genes and conditions (e.g., different samples). In the context of cancer, these structures are associated with significantly

---

This manuscript was submitted in IEEE Transaction on Knowledge and Data Engineering on 12/01/2017.
- J.-X. Liu is with the School of Information Science and Engineering, Qufu Normal University, Rizhao, Shandong, 276826 China (e-mail: sdcavell@126.com).
- C.-M. Feng (Corresponding author) is with the Bio-Computing Research Center, Shenzhen Graduate School, Harbin Institute of Technology, and the Key Laboratory of Network Oriented Intelligent Computation, Shenzhen, Guangdong, 518055 China; the School of Information Science and Engineering, Qufu Normal University, Rizhao, Shandong, 276826 China; (e-mail: fengchunmei0304@foxmail.com).
- X.-Z. Kong is with the School of Information Science and Engineering, Qufu Normal University, Rizhao, Shandong, 276826 China (e-mail: kongxzhen@163.com).
- Y. Xu (Corresponding author) is with the Bio-Computing Research Center, Shenzhen Graduate School, Harbin Institute of Technology, and with Key Laboratory of Network Oriented Intelligent Computation, Shenzhen, Guangdong, 518055 China (E-mail: yongxu@ymail.com).

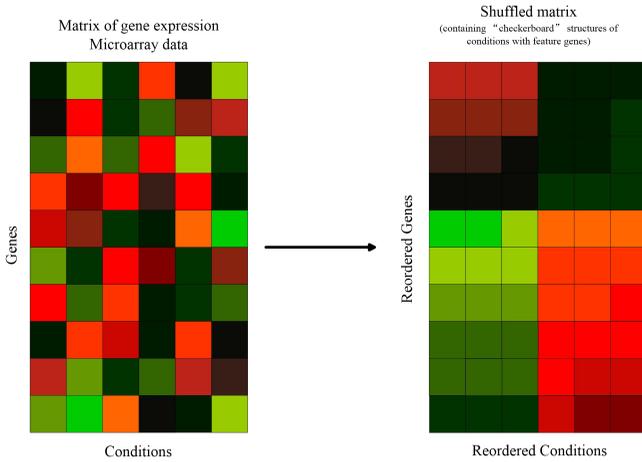

Fig. 1. Heat map of identifying checkerboard structures associated with feature genes and certain conditions. Left: the raw gene expression data, where each column corresponds to a sample. Right: the shuffled matrix containing checkerboard structures of conditions with feature genes.
Notes: All the original figures in this paper can be obtained in the separate file.

up-regulated or down-regulated genes in patients with specific types of tumors [8].

The raw gene expression data matrix is presented graphically in Fig. 1. In this matrix, the rows represent genes and the columns are different experimental conditions (e.g., different samples). Under this assumption, the matrix in Fig. 1 could be reorganized in a framework with a checkerboard-like structure. The various blocks in this structure are the strongly correlated genes (rows) over a subset of samples (columns). Medical researchers can develop personalized studies of different patients (samples) with a variety of regulatory genes.

PCA has been well applied in research and has yielded satisfactory results in clustering [9]. Although the features selected by PCA retain the main information of the original variables, and this information is the main part of the variation, there are weaknesses to this approach. For example, using limited data to obtain more useful information is one of the biggest bottlenecks. When PCA is applied for dimension reduction, it is essential to introduce manifold learning to learn the internal geometric structure. It has been proven that the incorporation of manifold learning into PCA facilitates the clustering effect [10]. Manifold learning finds the low-dimensional structure in high-dimensional data, which reveals the nonlinear geometric structure within the data [11]. The cross-application of manifold learning and other techniques has yielded satisfactory results [10, 12]. Nonlinear manifold learning algorithms include Laplacian Eigenmaps (LE), Isomap, and Locally Linear Embedding (LLE) [13-15]. For example, Jiang et al. proposed graph Laplacian PCA (gLPCA) and its robust model (RgLPCA), which introduced manifold learning into PCA [10]. Additional improved models based on gLPCA have been proposed, and good clustering and feature selection results have been obtained [16].

The methods mentioned above only focus on one-way clustering, which is solely based on genes or samples and ignores the relationship between them. Especially for high-dimensional, sparse and noisy data, it is difficult to meet the accuracy requirements in practice. As of now, there are more concerns about bi-clustering, which has achieved better results than one-way clustering. Bi-clustering clusters the rows and columns at the same time, which assist and restrict each other, and is also effective on high-dimensional and sparse data. In this paper, we incorporate manifold learning into the PCA model both in the principal directions (gene manifold) and along the PCs (condition manifold) to consider the internal structure of the data. In this way, the chessboard structures inside the observation data are constructed.

### 1.3 Uncovering checkerboard structures through dual graph-regularization PCA

Motivated by recent progress in the PCA method and bi-clustering [17, 18], we propose a novel method called dual graph-regularization principal component analysis (DGPCA). This method simultaneously considers the internal geometric structures of the condition manifold and the gene manifold. The geometric structures of the sample and gene spaces are encoded by constructing two nearest-neighbor graphs. To summarize, the main contributions of this paper are as follows:

1. We propose a novel PCA method named DGPCA. This method simultaneously considers the internal geometric structure information contained in both condition and gene data.
2. We present a closed-form solution for this problem and design an algorithm to address it, which avoids the instability of the iterative algorithm.
3. A visual checkerboard structure is found by the proposed method in combination with bi-clustering in the observed data. This structure corresponds to the genes that are significantly up-regulated or down-regulated in patients with certain types of tumors.
4. To detect new "marker genes" in the checkerboard structures, we mine the genes that are strongly regulated under the particular "conditions". These regulatory genes are more effective than graph-PCA-based methods. DGPCA provides a tool that is helpful for the study of the pathogenesis of cancer.

The rest of this paper is organized as follows. First, several related works are introduced in Section 2, including graph Laplacian PCA (gLPCA) and its robust models. Then, the method of DGPCA is first formulated in Section 3. The closed-form solution of this problem is also given in this section. Comprehensive experiments are carried out to evaluate the DGPCA method in Section 4. Finally, the paper is concluded in Section 5.

TABLE 1.
SOME NOTATIONS USED IN THIS PAPER

| Notation | Description |
| --- | --- |
| $\mathbf{X}$ | data matrix of size $d \times n$ |
| $n$ | number of samples |
| $d$ | number of genes |
| $k$ | number of clusters |
| $\mathbf{x}_i$ | $i$ th row of $\mathbf{X}$ |
| $\mathbf{x}_{:,i}$ | $i$ th column of $\mathbf{X}$ |
| $\mathbf{V}$ | the PCs of size $n \times k$ |
| $\mathbf{U}$ | the principal directions of size $d \times k$ |
| $\mathrm{Tr}(\mathbf{A})$ | the trace norm of the matrix $\mathbf{A}$ |
| $\|\mathbf{A}\|_F$ | the Frobenius norm of the matrix $\mathbf{A}$ |

## 2 RELATED WORK

Before we present the details of our method, some terms and notations are listed in Table 1, which will be frequently used in the following section. Then, we review some works that are related to this paper.

PCA finds the $k$-dimensional linear subspace where the projected data are as close as possible to the original data [10]. To provide an embedding for the data lying on a non-linear manifold, graph Laplacian PCA (gLPCA) is proposed [10]. The main task is to study the data matrix $\mathbf{X}$ that incorporates the cluster information into the graph data $\mathbf{W}$. This aim can be achieved by solving the following problem:

$$\min_{\mathbf{U},\mathbf{V}} \|\mathbf{X} - \mathbf{U}\mathbf{V}^T\|_F^2 + \gamma \mathrm{Tr}(\mathbf{V}^T \mathbf{L} \mathbf{V}) \quad s.t. \ \mathbf{V}^T \mathbf{V} = \mathbf{I}, \quad (1)$$

where $\gamma$ is the parameter that balances the contributions of the two terms; $\mathbf{L} = \mathbf{D} - \mathbf{W}$ is the graph Laplacian matrix, where $\mathbf{D}_{ii} = \sum_i \mathbf{W}_{ii}$ is a diagonal matrix whose elements are column or row sums of $\mathbf{W}$; and $\mathbf{W}$ is the weight matrix containing the edge weights of the graph with $n$ nodes. The definition of $\mathbf{W}_{ii}$ can be expressed as follows:

$$\mathbf{W}_{ij} = \begin{cases} 1 & if \ \mathbf{x}_i \in \mathbf{N}_k(\mathbf{x}_j) \ or \ \mathbf{x}_j \in \mathbf{N}_k(\mathbf{x}_i), \\ 0 & otherwise, \end{cases} \quad (2)$$

where $\mathbf{N}_k(\mathbf{x}_i)$ is the set of the $k$-nearest neighbors and edges connecting each data point $x_i$ in the graph.

The error function of gLPCA minimizes the sum of the squares of the data points that cause the data to deviate significantly from the model. To reduce the influence of outliers and noise, various robust versions of gLPCA have been proposed. These robust versions are formulated as follows:

$$\min_{\mathbf{U},\mathbf{V}} \|\mathbf{X} - \mathbf{U}\mathbf{V}^T\|_z + \gamma \mathrm{Tr}(\mathbf{V}^T \mathbf{L} \mathbf{V}) \quad s.t. \ \mathbf{V}^T \mathbf{V} = \mathbf{I}, \quad (3)$$

where $z$ can be the L$_{2,1}$-, L$_{1/2}$- or P-norm. The L$_{2,1}$-norm is defined as $\|\mathbf{A}\|_{2,1} = \sum_{j=1}^{n} \sqrt{\sum_{i=1}^{d} \mathbf{a}_{ij}^2} = \sum_{i=1}^{n} \|\mathbf{a}_i\|_2$ [10]; the L$_{1/2}$-norm is defined as $\|\mathbf{A}\|_{\frac{1}{2}}^{\frac{1}{2}} = \sum_{j=1}^{n} \sum_{i=1}^{d} |\mathbf{a}_{ij}|^{\frac{1}{2}}$ [19]; the P-norm is more flexible and can be tuned from 0 to 1 by a proximal operator: $shrink_p(\mathbf{t}, a) := \max\{0, |\mathbf{t}| - \alpha|\mathbf{t}|^{p-1}\}$, where $\mathbf{t}$ is a vector and $\alpha$ is the tuning parameter [20]. Good results can be obtained by these methods in feature selection and clustering. The robust versions utilize the L$_{2,1}$-, L$_{1/2}$- and P-norm in their error functions.

## 3 METHODOLOGY

### 3.1 Construct sample and gene graph

Recent research has shown that both the observed samples and genes lie on nonlinear low-dimensional manifolds, namely, the sample manifold and gene manifold, respectively [18]. Thus, we introduce two graphs to model the internal geometric structures of both the sample manifold and gene manifold. More specifically, we construct two graphs with different dimensions, namely the sample graph and gene graph, to explore the internal geometric structures of the rows and columns in gene expression data. The $k$-nearest-neighbor sample graph whose vertices correspond to $\{\mathbf{x}_{:,1},...,\mathbf{x}_{:,n}\}$ is first constructed. Following previous research, we use the 0-1 weighting scheme to construct the $k$-nearest-neighbor data graph [10]. The sample weight matrix can be defined as follows:

$$[\mathbf{W}_\mathbf{V}]_{ij} = \begin{cases} 1, & if \ \mathbf{x}_{:,j} \in N(\mathbf{x}_{:,i}), \\ 0, & otherwise, \end{cases} \quad i,j = 1,...,n, \quad (4)$$

where $N(\mathbf{x}_{:,i})$ is the $k$-nearest neighbor of $\mathbf{x}_{:,i}$ and $\mathbf{L}_\mathbf{V} = \mathbf{D}_\mathbf{V} - \mathbf{W}_\mathbf{V}$ is the Laplacian matrix of the sample, where $[\mathbf{D}_\mathbf{V}]_{ii} = \sum_j [\mathbf{W}_\mathbf{V}]_{ij}$ is a diagonal degree matrix. Similarly, we also use the 0-1 weighting scheme to construct the $k$-nearest-neighbor gene graph. $\{\mathbf{x}^T_{1,:},...,\mathbf{x}^T_{d,:}\}$ is the collection of vertices of the sample graph. The gene weight matrix is defined as follows:

$$[\mathbf{W}_\mathbf{U}]_{ij} = \begin{cases} 1, & if \ \mathbf{x}^T_{j,:} \in N(\mathbf{x}^T_{i,:}), \\ 0, & otherwise, \end{cases} \quad i,j = 1,...,d. \quad (5)$$

The gene graph Laplacian matrix is defined as $\mathbf{L}_\mathbf{U} = \mathbf{D}_\mathbf{U} - \mathbf{W}_\mathbf{U}$.

### 3.2 Objective function of dual graph-regularization PCA

Based on the graph regularizations of the sample manifold and gene manifold, we propose a new dual graph-regularization PCA method, with an objective function that is formulated as follows:

$$\min_{\mathbf{U},\mathbf{V}} \|\mathbf{X} - \mathbf{U}\mathbf{V}^T\|_F^2 + \alpha \mathrm{Tr}(\mathbf{U}^T \mathbf{L}_\mathbf{U} \mathbf{U}) + \beta \mathrm{Tr}(\mathbf{V}^T \mathbf{L}_\mathbf{V} \mathbf{V}) \quad (6)$$
$$s.t. \ \mathbf{V}^T \mathbf{V} = \mathbf{I},$$

where $\alpha$ and $\beta$ are parameters that balance the contributions from the reconstruction error of DGPCA in the first term and the graph regularizations in the latter two terms. When $\alpha = 0$, DGPCA degrades to the gLPCA method, and when $\alpha = \beta = 0$, DGPCA degrades to the standard PCA method.

### 3.3 Closed-form solution of DGPCA

We present a closed-form solution to the problem. The instability of the iterative solution can be avoided by our method. The objective function can be rewritten as follows:

$$\ell_{GDPCA} = \text{Tr}\left((\mathbf{X}-\mathbf{UV}^T)(\mathbf{X}-\mathbf{UV}^T)^T\right)$$
$$+\alpha \text{Tr}(\mathbf{U}^T\mathbf{L_U}\mathbf{U}) + \beta \text{Tr}(\mathbf{V}^T\mathbf{L_V}\mathbf{V}) \quad (7)$$
$$s.t. \ \mathbf{V}^T\mathbf{V}=\mathbf{I}.$$

First, by computing the optimal $\mathbf{U}$ while fixing $\mathbf{V}$, we can obtain the following results:

$$\frac{\partial \ell}{\partial \mathbf{U}} = -2(\mathbf{X}-\mathbf{UV}^T)\mathbf{V} + \alpha(\mathbf{L_U}\mathbf{U} + \mathbf{L_U}^T\mathbf{U}) = 0. \quad (8)$$

Thus, the optimal solution of $\mathbf{U}$ is given by
$$\mathbf{U} = \mathbf{A}^{-1}\mathbf{XV}. \quad (9)$$

Here, $\mathbf{A} = 1/2(2 + \alpha\mathbf{L_U} + \alpha\mathbf{L_U}^T)$. Then, we set $\mathbf{U} = \mathbf{A}^{-1}\mathbf{XV}$ and solve for the optimal $\mathbf{V}$. The objective function becomes

$$\min_{\mathbf{V}} \|\mathbf{X} - \mathbf{A}^{-1}\mathbf{XVV}^T\|_F^2 + \beta \text{Tr}(\mathbf{V}^T\mathbf{L_V}\mathbf{V}) \quad s.t. \ \mathbf{V}^T\mathbf{V}=\mathbf{I}. \quad (10)$$

By some algebra, we have

$$\text{Tr}(\mathbf{X}-\mathbf{A}^{-1}\mathbf{XVV}^T)(\mathbf{X}-\mathbf{A}^{-1}\mathbf{XVV}^T)^T + \beta \text{Tr}(\mathbf{V}^T\mathbf{L_V}\mathbf{V})$$
$$= \|\mathbf{X}\|_F^2 - 2\text{Tr}(\mathbf{XVV}^T\mathbf{X}(\mathbf{A}^{-1})^T)$$
$$+ \text{Tr}(\mathbf{A}^{-1}\mathbf{XVV}^T\mathbf{X}^T(\mathbf{A}^{-1})^T) + \beta \text{Tr}(\mathbf{V}^T\mathbf{L_V}\mathbf{V})$$
$$= \|\mathbf{X}\|_F^2 + \text{Tr}\left(\mathbf{V}^T\left(-2\mathbf{X}^T(\mathbf{A}^{-1})^T\mathbf{X} + \mathbf{X}^T(\mathbf{A}^{-1})^T\mathbf{A}^{-1}\mathbf{X}\right)\mathbf{V}\right) \quad (11)$$
$$+ \beta \text{Tr}(\mathbf{V}^T\mathbf{L_V}\mathbf{V})$$
$$= \|\mathbf{X}\|_F^2 + \text{Tr}\left(\mathbf{V}^T\begin{pmatrix}-2\mathbf{X}^T(\mathbf{A}^{-1})^T\mathbf{X} \\ +\mathbf{X}^T(\mathbf{A}^{-1})^T\mathbf{A}^{-1}\mathbf{X} + \beta\mathbf{L_V}\end{pmatrix}\mathbf{V}\right).$$

As a result, Eq. (10) is equivalent to the following:
$$\ell(\mathbf{V}) = \min_{\mathbf{V}} \text{Tr}(\mathbf{V}^T\mathbf{B}\mathbf{V}) \quad s.t. \ \mathbf{V}^T\mathbf{V}=\mathbf{I}, \quad (12)$$

where $\mathbf{B} = -2\mathbf{X}^T(\mathbf{A}^{-1})^T\mathbf{X} + \mathbf{X}^T(\mathbf{A}^{-1})^T\mathbf{A}^{-1}\mathbf{X} + \beta\mathbf{L_V}$.

Therefore, the optimization problem can be solved by the eigenvectors corresponding to the $k$ smallest eigenvalues of the matrix $\mathbf{B}$.

## 4 EXPERIMENTS

The primary goal of the experiments is to evaluate the proposed DGPCA method in comparison with gLPCA because DGPCA incorporates the gene manifold based on gLPCA. For completeness, we also compare our results to existing research results and the results of some other graph-Laplacian-PCA-based methods, such as RgLPCA [10], $L_{1/2}$gLPCA [19] and PgLPCA [20]. These methods focus on one-way clustering and only learn the sample geometry with the PCA method. First, we present a visual heat-map to display the results of bi-clustering to find "checkerboard" structures, if they exist. Then, experiments on selecting regulatory genes are presented to evaluate the performance of DGPCA compared with other methods and the existing research results. Biological analysis of these genes provides the basis for further research on new cancer markers. The experimental datasets and the parameter settings for each method are described in the following subsections.

### 4.1 Datasets

The datasets used in these experiments are described as follows:

Leukemia data: The leukemia data consist of a matrix that includes 38 samples and 5000 genes. This dataset is publicly available at https://sites.google.com/site/feipingnie/file. It contains 11 types of acute myelogenous leukemia (AML) and 27 types of acute lymphoblastic leukemia (ALL), and ALL is divided into T- and B-cell subtypes [21].

Colon cancer data: The colon cancer data consists of a matrix that includes 2000 genes and 62 tissues. These tissues are divided into 22 normal and 40 colon tumor samples [22]. This dataset and its detailed description are publicly available at http://genomics-pubs.princeton.edu/oncology/affydata/index.html

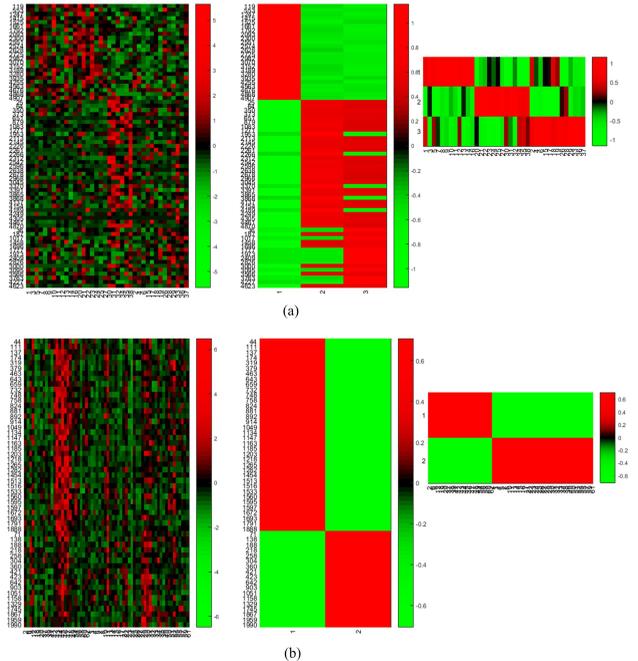

Fig. 2. Heat map of DGPCA bi-clustering result on leukemia (a) and colon cancer data (b). Left: the checkerboard structure of the leukemia data, where each column corresponds to a sample. Center: the principal directions (gene manifold). Right: the projected samples in the new subspace (condition manifold).

Notes: All the original figures in this paper can be obtained in the separate file.

## 4.2 Experimental setting

For each method, all the parameters are tuned to search for the optimal value. Since the parameter $\gamma$ on gLPCA, RgLPCA, and L$_{1/2}$gLPCA, and the parameter P on PgLPCA can be tuned in the range of 0~1, we search for their optimal values in $[0:0.1:1]$. Within the given range, the greater the value of $\gamma$, the greater the role of the graph Laplacian in the objective function. According to previous research, we set $\gamma = 0.5$ to obtain fair results [10, 19, 20]. In practice, we set the parameter $\rho = 1.2$ in RgLPCA and L$_{1/2}$gLPCA. A wider range has also been investigated, but $\rho = 1.1 \sim 1.5$ yields good results.

Since the parameters $\alpha$ and $\beta$ of the proposed method have no special limits, we search for their optimal values in $\{50^{-4}, 50^{-2}, 50^0, 50^2, 50^4\}$. We set $\beta = \gamma = 0.5$ to obtain the condition manifold because both the parameters $\beta$ and $\gamma$ control the contribution of the graph Laplacian in the samples. In practice, we find that when $\alpha = 0.05$, satisfactory results can be obtained. Based on the number of categories of the experimental data, we set the numbers of reduced dimensions $k_1 = 3$ and $k_2 = 2$ for leukemia data and colon cancer data, respectively.

## 4.3 Bi-clustering results to find "checkerboard" structure

Our proposed method provides a geometric structure of not only the condition manifold but also the gene manifold. To assess our method, it is useful to observe how well it performs on several gene expression datasets, with respect to achieving the goal of finding checkerboard structures. Since the previous PCA-based methods ignore the joint geometric structures of conditions and genes, they are not designed to reveal the checkerboard structures of gene expression data. Therefore, only the proposed method DGPCA is evaluated through bi-clustering and the visual heatmap. We use bi-clustering as a tool for data visualization and reasonable interpretation of our method. Our method employs manifold learning schemes that highlight the internal geometric structures of both genes and conditions, thereby directly revealing the degree of bi-clustering.

We apply the proposed method to two publicly available datasets: leukemia data and colon cancer data. The visual heatmap is used to display the results of bi-clustering to find checkerboard structures, if they do exist. The heat maps of (a) and (b) in Fig. 2 display the bi-clustering results of DGPCA on the leukemia and colon cancer data, respectively. In this figure, on the left is the checkerboard structure of the leukemia data, where each column corresponds to a sample; in the center are the principal directions (gene manifold); and on the right are the projected samples in the new subspace (condition manifold). The two coordinates represent the sample number and gene expression level. From Fig. 2, we can observe that the two raw datasets are rearranged in a checkerboard structure. In heat map (a), the arrangement of the 38 samples is generally based on the three types of labels: AML, T- and B-cells. A similar conclusion can be drawn from the colon data in heat map (b). The 62 tissues are generally arranged according to two types of labels because these tissues are divided into normal and colon tumor samples. The cross-section of the different colors shows the interaction between the samples and the genes. Specifically, blocks of different colors represent the clustering results of different data classes under the interaction of the gene and condition manifolds. These consistently formatted graphs show the checkerboard structure of each data class on the condition manifold

TABLE 2.
RESULTS ON TOTAL RELEVANCE SCORES (TRS) OF GLPCA, RGLPCA, L1/2GLPCA AND DGPCA. THE BEST RESULTS ARE HIGHLIGHTED IN BOLD.

| Data set | leukemia data | | | colon cancer data | | |
|---|---|---|---|---|---|---|
| | TRS | ACC | ARS | TRS | ACC | ARS |
| PCA | 306.19 | 53.00 | 5.78 | 134.11 | 22.00 | 6.10 |
| gLPCA | 324.67 | 55.00 | 5.90 | 126.87 | 22.00 | 5.77 |
| RgLPCA | 306.19 | 53.00 | 5.78 | 134.11 | 22.00 | 6.10 |
| L$_{1/2}$gLPCA | 306.19 | 53.00 | 5.78 | 134.11 | 22.00 | 6.10 |
| PgLPCA | 384.42 | **58.00** | 5.77 | 147.51 | 26.00 | 5.67 |
| DGPCA | **395.98** | 52.00 | **7.62** | **243.98** | **27.00** | **9.04** |

TABLE 3.
THE DETAILED INFORMATION OF THE PECULIAR GENES ON LEUKEMIA DATA SELECTED BY OUR METHOD, INCLUDING GENE FUNCTION, ASSOCIATED DISEASE AND RELEVANCE SCORE FOR LEUKEMIA. THIS TABLE LISTS ONLY GENES WITH RELEVANCE SCORES GREATER THAN 10.

| Gene | Gene function | Relevance score |
|---|---|---|
| FLT3 | This gene encodes a class III receptor tyrosine kinase that regulates hematopoiesis. | 79.21 |
| MYC | This functions as a transcription factor that regulates transcription of specific target genes. | 42.66 |
| MPO | This is a heme protein synthesized during myeloid differentiation that constitutes the major component of neutrophil azurophilic granules. | 19.06 |
| CCNA1 | The protein encoded by this gene belongs to the highly conserved cyclin family, in which members are characterized by a dramatic periodicity in protein abundance throughout the cell cycle. | 14.29 |
| MS4A1 | This gene encodes a member of the membrane-spanning 4A gene family. | 10.65 |
| CALR | This gene acts as an important modulator of the regulation of gene transcription by nuclear hormone receptors. | 10.1 |

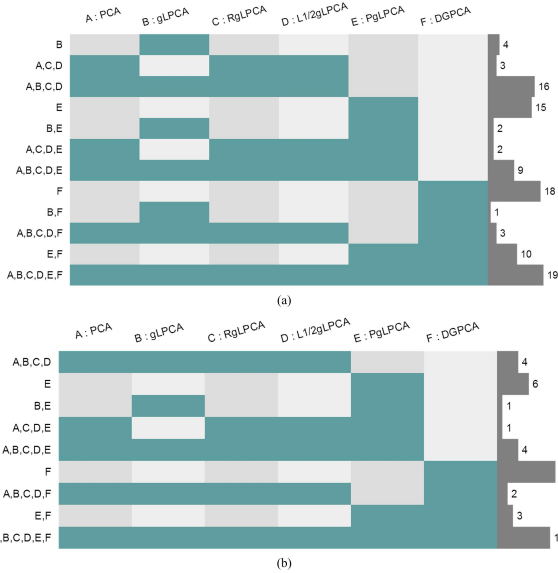

(a)

(b)

Fig. 3. Overlap among the differentially expressed genes identified by the compared methods.
Notes: All the original figures in this paper can be obtained in the separate file.

together with the gene manifold.

## 4.4 Finding regulatory genes under the particular "conditions"

Since bi-clustering is used as a tool for data visualization and interpretation, it is natural to detect the quality of bi-clusters in terms of biological significance or accuracy. Here, we perform a study to assess the quality of our method, in which we apply DGPCA to the experimental data used in this paper. The top-100 regulatory genes are selected from both the leukemia and colon cancer data for analysis. First, we rank the scores of all genes in descending order. Then, the regulatory genes can be extracted by the corresponding indices. In other words, the extracted top genes with high scores can be deemed regulatory genes. DAVID 6.7 is used as a tool to find the official names of the selected regulatory genes from the leukemia data; it is publicly available at https://david-d.ncifcrf.gov/. For the colon cancer data, we search for the abbreviation of each gene using ToppGene Suite, which is publicly available at https://toppgene.cchmc.org/enrichment.jsp. We download the pathogenic gene pools for leukemia and colon cancer from GeneCards. This searchable, integrative database is publicly available at http://www.genecards.org/. Assume the selected genes match the pathogenic gene pool over the two experimental datasets.

### 4.4.1 Analysis of matching results

The matching results on the leukemia and colon cancer data of gLPCA, RgLPCA, $L_{1/2}$gLPCA, PgLPCA and DGPCA are listed in a separate file. In this file, regulatory genes are selected by all compared methods, where the marked genes denote peculiar pathogenic genes selected by our method but not by other methods. The relative scores of each regulatory gene associated with the disease are also listed in this file. To determine the efficiency of the identified regulatory genes, we summarize the total relevance scores (TRS), accuracies (ACC) and average relevance scores (ARS) in Table 2. The best results are highlighted in bold. ACC is the accuracy of the regulatory genes from the selected top-100 genes; it is defined as follows:

$$\text{ACC} = \frac{\sum_{i=1}^{n}\delta(p_i, map(q_i))}{n} \times 100\%, \quad (13)$$

where $q_i$ is a regulatory gene that was selected by our method, and $p_i$ is a pathogenic gene of the disease. $\delta(p_i, map(q_i))$ is given by

$$\delta(x,y) = \begin{cases} 1, & x = y, \\ 0, & otherwise, \end{cases} \quad (14)$$

where $map(q_i)$ is the mapping function. A higher ACC values indicates improved performance. From these tables, we make the following observations:

1. The lowest TRS are obtained by PCA, on both the leukemia and colon cancer data. This is reasonable, since classical PCA is not robust enough.
2. By considering the internal geometric structures, gLPCA achieves some improvement over classical PCA.
3. RgLPCA, $L_{1/2}$gLPCA and PgLPCA also aim at improving the robustness of the algorithm. PgLPCA outperforms the others because this method provides the utmost flexibility, since the value of P can be tuned from 0 to 1.
4. The ACC result of DGPCA on the colon cancer data is the highest and that on the leukemia data is the lowest. The large amount of noise in the leukemia data might be the main cause, which leads to the ACC result of DGPCA being less than those of robust methods such as RgLPCA, $L_{1/2}$gLPCA and PgLPCA. These methods are designed to improve the robustness and reduce the effects of noise and outliers. However, the TRS and ARS values obtained by DGPCA are the highest on both datasets because DGPCA picked out some important genes that were ignored by the other methods.

TABLE 4.
THE DETAILED INFORMATION OF THE PECULIAR GENES ON COLON CANCER DATA SELECTED BY OUR METHOD, INCLUDING GENE FUNCTION AND RELEVANCE SCORE WITH COLON CANCER. THIS TABLE LISTS ONLY GENES WITH RELEVANCE SCORES GREATER THAN 10.

| Gene | Gene function | Relevance score |
|---|---|---|
| GSTM1 | This gene encodes a glutathione S-transferase that belongs to the mu class. | 49.89 |
| SRC | The protein encoded by this gene is a tyrosine-protein kinase whose activity can be inhibited by phosphorylation by c-SRC kinase. | 47.03 |

### 4.4.2 Visualization of overlapping results

Here, we utilize a Venn diagram to visualize the overlap among the regulatory genes selected by the compared methods. It was obtained using OmicsBean, which is a multi-omics data analysis tool, which is available at http://www.omicsbean.com:88/. The Venn diagram in Fig. 3 shows the following overlapping results: (a) is the overlapping result on the leukemia data and (b) is the overlapping result on the colon cancer data. The different permutations and combinations of all methods are displayed in the left coordinate of the Venn diagram. When there is only one method on the left coordinate, the corresponding number on the right represents a regulatory gene that is obtained only by this method. When there are several methods on the left coordinate, the corresponding number on the right indicates are regulatory gene that is obtained by these methods.

As shown in Fig. 3, the proposed method selects the largest number of regulatory genes that are not selected by other methods. The numbers of such unique regulatory genes from the two datasets that were excavated by the proposed method are 18 and 11. According to the relevance scores in the additional file, these genes are highly related to disease and cannot be neglected in the study of leukemia and colon cancer. In contrast, few unique regulatory genes are excavated by PgLPCA and gLPCA. Since the relevance scores of these genes with respect to disease are not high, no further research on them will be conducted in this paper.

### 4.4.3 Comparison with published results

To evaluate the promising results of DGPCA, we compared all related genes, such as biomarker or characteristic genes, that were obtained by other methods in the literature, to the regulatory genes found in this paper. We found that 20 of the 30 genes identified by the P-norm Robust Feature Extraction (PRFE) method are related to leukemia and the ARS is 6.56 [23]. Liu et al. identified the feature genes from leukemia data by combining RPCA and LDA [24]. Eleven genes were identified as characteristic genes associated with leukemia, and the highest relevance score was 19.06. The average and highest scores of the characteristic genes selected by RGNMF from the leukemia data were 7.31 and 25.99, respectively[25]. Among all the biomarkers selected by Wu et al., 29 out of 52 genes selected by our method can be found in this article [26]. However, the 23 genes they ignored include several important pathogenic genes, which provide evidence that our method outperforms the method in the previous study. In particular, some important genes with high relevance scores were not excavated in this paper. Some researchers have made outstanding contributions to the discovery of colon-cancer-associated genes, but most of the regulatory genes found in this paper have been neglected [27, 28]. These genes are considered to be new oncogenes and have research value for leukemia and colon cancer. Further studies of these genes are conducted in the next subsection.

## 4.5 Function analysis of unique regulatory genes

Some genes acquired by DGPCA that have been ignored by the existing methods are important contributions to the study of related cancers. Thus, further analysis of these genes is necessary [26]. We list detailed information on the unique regulatory genes with relevance scores greater than 10 in Table 3 and Table 4. These genes will facilitate the study of leukemia and colon cancer in clinical practice. The functions and relevance scores of these genes have been summarized. Some of the results from these tables and from other research are summarized as follows:

1. Far more unique regulatory genes are identified by DGPCA from the leukemia data than from the colon cancer data. This large disparity is due to the distinct attributes of different datasets. For example, the labels of the leukemia data are divided into three types, whereas those of the colon cancer data are divided into two types.
2. In Table 3, FLT3 has the highest relevance score with respect to leukemia in its pathogenic gene pool, and other methods do not identify this gene. Various published articles have studied the relationship between FLT3 and leukemia, which indicates that it is an important gene in leukemia research [29, 30]. FLT3 is a gene that cannot be ignored in the study of leukemia, and it highlights the accuracy of our method. The relevance score of MYC with respect to leukemia is 42.66. GeneCards indicates that acute lymphocytic leukemia, acute lymphoblastic 3 and Burkitt lymphoma are the diseases associated with this gene. Neither of these genes can be ignored in the study of leukemia, and they are not found by other methods. Other genes in Table 3 also have certain degrees of correlation with leukemia [31, 32]. These important genes are missing from the 210 leukemia-related biomarkers identified by Wu et al. [26]. It is obvious that our results better enhance the clinical studies of leukemia, compared to those of other methods. The high relevance scores reflect the close relationships of CCNA1 and MS4A1 with leukemia. These results provide a great research space for us to study these genes, because there is little biological research on this topic.
3. In Table 4, the relevance scores of GSTM1 and SRC with respect to leukemia are 49.89 and 47.03, respectively. Many investigators have conducted studies on the regulation of these two genes in colon cancer [33]. Diseases associated with these genes include colon cancer and lung cancer. It has been proven in published articles that there is a certain relationship between them [34, 35]. The above results demonstrate that our method achieves a more accurate performance than earlier methods.

## 4.6 Gene interaction of biological pathway analysis

In addition to analyzing the selected regulatory genes for assessing the quality of bi-clusters, it is natural to relate those genes together with conditions to biological pathways. We send these selected regulatory genes to KEGG (http://www.kegg.jp/) to find the biological pathways. KEGG is a biological database resource for realizing high-level functions and applying biological systems. This database provides new perspectives on genomes, biological pathways, diseases and drugs [36]. The pathway graph of two datasets can be found in separate file, where the genes in pink are human disease genes, genes in blue are drug target genes and green are human genes.

The biological pathways of the hematopoietic cell lineage record the change process of hematopoietic stem cells (HSC). These cells can undergo self-renewal or differentiation into a common lymphoid progenitor (CLP) or a common myeloid progenitor (CMP). A CLP gives rise to the lymphoid lineage of white blood cells or the natural killer (NK) cells of leukocytes on the T and B lymphocytes. This process increases the production of platelets and clotting. Therefore, cellular stages are determined by the specific expression states of these genes. The interaction of genes selected by our method precisely describes the work of the hematopoietic system and is also the major cause of leukemia. The pathways of ribosome process and translate genetic information. Kimura *et al*. conducted studies on the ribosome and colon cancer cell lines to demonstrate the role of the ribosome in some biological processes [37]. The above two biological pathways reflect the interaction of genes in the corresponding datasets.

## 5 CONCLUSION

Following the idea of bi-clustering and the relevance of rows and columns based on gene expression data, this paper presented a novel method called DGPCA. This method incorporated the information obtained by the gene manifold to improve the clustering of conditions in the model. In particular, the gene and condition manifolds could be simultaneously obtained by gene clusters and tumor clusters. The visual heatmap displayed the results of bi-clustering to find "checkerboard" structures. In situations where the checkerboard structure is found, the regulatory genes are selected compared with those obtained by other PCA-based methods and those in published articles to evaluate the quality of the bi-clusters. The identified regulatory genes have been analyzed in terms of function and co-expression (pathways). DGPCA provides a tool that is helpful for the study of the pathogenesis of cancer.

## ACKNOWLEDGMENT

This work was supported in part by the NSFC under grant Nos. 61572284, 61502272 and 61702299. C.-M. Feng and Y. Xu are the corresponding authors.

## REFERENCES

[1] R. A. Irizarry, Z. Wu, and H. A. Jaffee, "Comparison of Affymetrix GeneChip expression measures," *Bioinformatics*, vol. 22, no. 7, pp. 789-94, 2006.

[2] J. X. Liu, D. Wang, Y. L. Gao, C. H. Zheng, J. L. Shang, F. Liu, and Y. Xu, "A Joint-L 2,1 -norm-Constraint-based Semi-Supervised Feature Extraction for RNA-Seq Data Analysis," *Neurocomputing,* vol. 228, pp. 263-269, 8 March 2017.

[3] J. X. Liu, Y. Xu, Y. L. Gao, C. H. Zheng, D. Wang, and Q. Zhu, "A Class-Information-Based Sparse Component Analysis Method to Identify Differentially Expressed Genes on RNA-Seq Data," *IEEE/ACM Transactions on Computational Biology & Bioinformatics*, vol. 13, no. 2, pp. 392-398, 2015.

[4] D. Hanisch, A. Zien, R. Zimmer, and T. Lengauer, "Co-clustering of biological networks and gene expression data," *Bioinformatics,* vol. 18 Suppl 1, no. suppl_1, pp. S145-54, 2002.

[5] F. Shang, L. C. Jiao, and F. Wang, "Graph dual regularization non-negative matrix factorization for co-clustering," *Pattern Recognition*, vol. 45, no. 6, pp. 2237-2250, 2012.

[6] G. Getz, E. Levine, and E. Domany, "Coupled Two-Way Clustering Analysis of Gene Microarray Data," *Proceedings of the National Academy of Sciences of the United States of America,* vol. 97, no. 22, pp. 12079, 2000.

[7] S. C. Madeira, and A. L. Oliveira, "Biclustering algorithms for biological data analysis: a survey," *IEEE/ACM Transactions on Computational Biology & Bioinformatics,* vol. 1, no. 1, pp. 24-45, 2004.

[8] Y. Kluger, R. Basri, J. T. Chang, and M. Gerstein, "Spectral Biclustering of Microarray Data: Coclustering Genes and Conditions," *Genome Research,* vol. 13, no. 4, pp. 703-716, 2003.

[9] K. Y. Yeung, and W. L. Ruzzo, "Principal component analysis for clustering gene expression data," *Bioinformatics,* vol. 17, no. 9, pp. 763-774, 2001.

[10] B. Jiang, C. Ding, and J. Tang, "Graph-laplacian pca: Closed-form solution and robustness." *Computer Vision & Pattern Recognition* pp. 3492-3498.

[11] B. Raducanu, and F. Dornaika, "A supervised non-linear dimensionality reduction approach for manifold learning," *Pattern Recognition,* vol. 45, no. 6, pp. 2432-2444, 2012.

[12] D. Cai, X. He, J. Han, and T. S. Huang, "Graph regularized nonnegative matrix factorization for data representation," *IEEE Transactions on Pattern Analysis and Machine Intelligence,* vol. 33, no. 8, pp. 1548-1560, 2011.

[13] M. Belkin, and P. Niyogi, "Laplacian Eigenmaps for dimensionality reduction and data representation," *Neural Computation,* vol. 15, no. 15, pp. 1373-1396, 2003.

[14] M. Balasubramanian, and E. L. Schwartz, "The isomap algorithm and topological stability," *Science,* vol. 295, no. 5552, pp. 7, 2002.

[15] S. T. Roweis, and L. K. Saul, "Nonlinear dimensionality reduction by locally linear embedding," *Science,* vol. 290, no. 5500, pp. 2323, 2000.

[16] C. M. Feng, J. X. Liu, Y. L. Gao, J. Wang, D. Q. Wang, and Y. Du, "A graph-Laplacian PCA based on L1/2-norm constraint for characteristic gene selection," *IEEE International Conference on Bioinformatics and Biomedicine (BIBM),* vol. 10, pp. 1795-1799, 2016.

[17] V. Sindhwani, J. Hu, and A. Mojsilovic, "Regularized Co-Clustering with Dual Supervision." *International Conference on Neural Information Processing Systems* pp. 1505-1512.

[18] Q. Gu, and J. Zhou, "Co-clustering on manifolds." *Acm Sigkdd International Conference on Knowledge Discovery & Data Mining*, pp. 359-368.

[19] C. M. Feng, Y. L. Gao, J. X. Liu, J. Wang, C. G. Wen, and D. Q. Wang, "JointL1/2-Norm Constraint and Graph-Laplacian PCA Method for Feature Extraction," *BioMed Research International*,2017,(2017-4-2), vol. 2017, pp. 1-14, 2017.

[20] C. M. Feng, Y. L. Gao, J. X. Liu, C. H. Zheng, and J. Yu, "PCA based on Graph Laplacian Regularization and P-norm for Gene Selection and Clustering," *IEEE Transactions on Nanobioscience*, vol. 16, no. 4, pp. 257-265, 2017.

[21] J.-P. Brunet, P. Tamayo, T. R. Golub, and J. P. Mesirov, "Metagenes and molecular pattern discovery using matrix factorization," *Proceedings of the national academy of sciences*, vol. 101, no. 12, pp. 4164-4169, 2004.

[22] G. Getz, H. Gal, I. Kela, D. A. Notterman, and E. Domany, "Coupled two-way clustering analysis of breast cancer and colon cancer gene expression data," *Bioinformatics*, vol. 19, no. 9, pp. 1079-1089, 2003.

[23] J. Liu, J. X. Liu, Y. L. Gao, X. Z. Kong, X. S. Wang, and D. Wang, "A P-Norm Robust Feature Extraction Method for Identifying Differentially Expressed Genes," *Plos One*, vol. 10, no. 7, 2015.

[24] J. X. Liu, Y. Xu, C. H. Zheng, H. Kong, and Z. H. Lai, "RPCA-Based Tumor Classification Using Gene Expression Data," *IEEE/ACM Transactions on Computational Biology & Bioinformatics*, vol. 12, no. 4, pp. 964-970, 2015.

[25] D. Wang, J. X. Liu, Y. L. Gao, C. H. Zheng, and Y. Xu, "Characteristic Gene Selection based on Robust Graph Regularized Non-negative Matrix Factorization," *IEEE/ACM Transactions on Computational Biology & Bioinformatics*, vol. 13, no. 6, pp. 1059-1067, 2016.

[26] M.-Y. Wu, D.-Q. Dai, X.-F. Zhang, and Y. Zhu, "Cancer subtype discovery and biomarker identification via a New robust network clustering algorithm," *PloS one*, vol. 8, no. 6, pp. e66256, 2013.

[27] C. C. Morgan, K. Shakya, A. Webb, T. A. Walsh, M. Lynch, C. E. Loscher, H. J. Ruskin, and M. J. O'Connell, "Colon cancer associated genes exhibit signatures of positive selection at functionally significant positions," *Bmc Evolutionary Biology*, vol. 12, no. 1, pp. 114, 2012.

[28] A. C. Goulet, G. Watts, J. L. Lord, and M. A. Nelson, "Profiling of selenomethionine responsive genes in colon cancer by microarray analysis," *Cancer Biology & Therapy*, vol. 6, no. 4, pp 494-503, 2007.

[29] D. G. Gilliland, and J. D. Griffin, "The roles of FLT3 in hematopoiesis and leukemia," *Blood*, vol. 100, no. 5, pp. 1532-42, 2002.

[30] M. Levis, and D. Small, "FLT3: ITDoes matter in leukemia," *Leukemia*, vol. 17, no. 9, pp. 1738, 2003.

[31] M. F. Jia, Y. M. Xi, X. E. Shi, H. Zhang, W. Deng, M. Li, P. Li, J. W. Xu, H. Z. Ma, and X. J. Yao, "[Relationship of MPO and NQO1 gene polymorphisms with susceptibility to acute leukemia]," *Journal of Experimental Hematology*, vol. 20, no. 6, pp. 1336-1340, 2012.

[32] I. Bonzheim, B. Mankel, P. Klapthor, J. Schmidt, T. Hinrichsen, O. Wachter, F. Fend, and L. Quintanillamartinez, "CALR-mutated essential thrombocythemia evolving to chronic myeloid leukemia with coexistent CALR mutation and BCR-ABL translocation," *Blood*, vol. 125, no. 14, pp. 2309-2311, 2015.

[33] M. L. Slattery, J. D. Potter, K. N. Ma, B. J. Caan, M. Leppert, and W. Samowitz, "Western diet, family history of colorectal cancer, NAT2, GSTM-1 and risk of colon cancer," *Cancer Causes & Control Ccc*, vol. 11, no. 1, pp. 1-8, 2000.

[34] A. Zupa, A. Sgambato, G. Bianchino, G. Improta, V. Grieco, G. L. Torre, B. Campisi, A. Traficante, M. Aieta, and A. Cittadini, "GSTM1 and NAT2 polymorphisms and colon, lung and bladder cancer risk: a case-control study," *Anticancer Research*, vol. 29, no. 5, pp. 1709, 2009.

[35] T. C. Windham, D. R. Parikh NUSiwak, J. M. Summy, D. J. Mcconkey, A. J. Kraker, and G. E. Gallick, "Src activation regulates anoikis in human colon tumor cell lines," *Oncogene*, vol. 21, no. 51, pp. 7797-7807, 2002.

[36] M. Kanehisa, M. Furumichi, T. Mao, Y. Sato, and K. Morishima, "KEGG: new perspectives on genomes, pathways, diseases and drugs," *Nucleic Acids Research*, vol. 45, no. D1, pp. D353, 2017.

[37] K. Kimura, A. Wada, M. Ueta, A. Ogata, S. Tanaka, A. Sakai, H. Yoshida, H. Fushitani, A. Miyamoto, and M. Fukushima, "Comparative proteomic analysis of the ribosomes in 5-fluorouracil resistance of a human colon cancer cell line using the radical-free and highly reducing method of two-dimensional polyacrylamide gel electrophoresis," *International Journal of Oncology*, vol. 37, no. 5, pp. 1271, 2010.

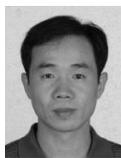

**Jin-Xing Liu** received a B.S. degree in electronic information and electrical engineering from Shandong University, China, in 1997; an M.S. degree in control theory and control engineering from QuFu Normal University, China, in 2003; and a Ph.D. degree in computer simulation and control from the South China University of Technology, China, in 2008. He is a professor at the School of Information Science and Engineering, Qufu Normal University, Rizhao, China. His research interests include pattern recognition, and bioinformatics.

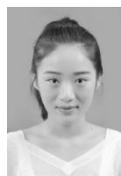

**Chun-Mei Feng** is a master's degree candidate in the Computer Science and Technology College of QuFu Normal University, China. She is currently pursuing the Ph.D. degree in computer science and technology at Harbin Institute of Technology, Shenzhen. Her research interests include feature selection, pattern recognition, bioinformatics and deep learning.

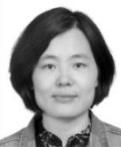

**Xiang-Zhen Kong** received a B.S. degree in computer science and technology from Qufu Normal University, China, in 2002 and an M.S. degree in control theory and control engineering from QuFu Normal University, China, in 2008. She is a lecturer at the School of Information Science and Engineering, Qufu Normal University. Her research interests include pattern recognition and bioinformatics.

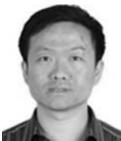

**Yong Xu** received his B.S. and M.S. degrees at the Air Force Institute of Meteorology (China) in 1994 and 1997, respectively. He then received his Ph.D. degree in pattern recognition and intelligence systems at the Nanjing University of Science and Technology in 2005. From May 2005 to April 2007, he worked at Shenzhen Graduate School, Harbin Institute of Technology as a postdoctoral research fellow. Now he is a professor at Shenzhen Graduate School, HIT. His current interests include pattern recognition, machine learning, and bioinformatics.